\documentclass[preprint]{aastex63}

\shortauthors{Sharkey et al.}
\graphicspath{{./}{figures/}}

\begin{document}

\title{Complex Water Ice Mixtures on NII Nereid: Constraints from NIR Reflectance}

\correspondingauthor{Benjamin N. L. Sharkey}
\email{sharkey@lpl.arizona.edu}

\author{Benjamin N. L. Sharkey}
\altaffiliation{Visiting Astronomer at the Infrared Telescope Facility, which is operated by the University of Hawaii under contract NNH14CK55B with the National Aeronautics and Space Administration.}
\affiliation{Lunar and Planetary Laboratory \\
University of Arizona \\
1629 E University Blvd \\
Tucson AZ 85721-0092, USA}

\author[0000-0002-7743-3491]{Vishnu Reddy}
\affiliation{Lunar and Planetary Laboratory \\
University of Arizona \\
1629 E University Blvd \\
Tucson AZ 85721-0092, USA}

\author{Juan A. Sanchez}
\affiliation{Planetary Science Institute \\
1700 E Fort Lowell Suite 106 \\
Tucson, AZ, 85719, USA}

\author[0000-0001-5456-2912]{Matthew R. M. Izawa}
\affiliation{Institute for Planetary Materials\\
Okayama University \\
827 Yamada, Misasa, Tottori 682-0193, Japan}

\author{Walter M. Harris}
\affiliation{Lunar and Planetary Laboratory \\
University of Arizona \\
1629 E University Blvd \\
Tucson AZ 85721-0092, USA}

\begin{abstract}
Nereid, Neptune's third largest satellite, lies in an irregular orbit and is the only outer satellite in the system (apart from Triton) that can be spectroscopically characterized with the current generation of Earth-based telescopes. We report our results on spectral characterization of Nereid using its reflectance spectrum from 0.8-2.4 $\mu m$, providing the first measurements over the range of 0.8-1.4 $\mu m$. We detect spectral absorption features of crystalline water ice in close agreement with previous measurements. We show that model fits of simple intimate mixtures including water ice do not provide simultaneous matches to absorption band depths at 1.5 and 2.0 $\mu m$ when accounting for the spectral continuum. Possible solutions include invoking a more complex continuum, including both crystalline and amorphous water ice, and allowing for sub-micron sized grains. We show that mixtures including magnetite and the CM2 chondrite Murchison provide a flexible framework for interpreting spectral variation of bodies with neutral-sloped spectra like that of Nereid. Magnetite in particular provides a good match to the spectral continuum without requiring the presence of Tholin-like organics. We note that carbonaceous chondrites and their components may be useful analogs for the non-ice components of outer solar system bodies, consistent with recent findings by Fraser et al. (2019). Comparison to spectra of large TNOs and satellites of Uranus show that Nereid's low albedo, deep water bands, and neutral color is distinct from many other icy objects, but such comparisons are limited by incomplete understanding of spectral variability among $\sim$100km-sized icy bodies. 

\end{abstract}



\section{Introduction} \label{sec:intro}
The Neptunian satellite system stands unique among those of the giant planets of the solar system. Neptune lacks any large, regular prograde satellites that can be understood to have formed concurrently with the planet (i.e. in a circumplanetary gas disk like the Galilean moons of Jupiter). Apart from Triton, no other satellite is over 500 km in diameter, and Triton’s retrograde orbit is suggestive of capture from a heliocentric orbit instead of co-accretion with Neptune \citep[e.g.,][]{mckinnon1984,agnor2006}. Nereid, \citep[340 $\pm$ 50 km diameter,][]{thomas1991}, the third largest satellite behind Triton and the inner satellite Proteus \citep[418 $\pm$ 16 km diameter,][]{Croft1992},  is also in a retrograde orbit and is exterior to Triton.

Recent modeling has suggested that the current Neptunian system can be replicated from an initial satellite system (with a mass distribution similar to the Uranian system) disrupted by the capture of Triton \citep{rufu2017}. Additionally, this model provides a pathway that allows for Triton’s capture to occur after the formation or capture of Nereid, the largest satellite exterior to Triton. The survival of Nereid-type satellites in this scenario contrasts with other hypothesized paths for the history of the Neptunian system, which argue that the small irregular satellites were emplaced after Triton \citep[as reviewed in][]{nogueira2011}.

The reflectance properties of Nereid were previously observed from 1.4-2.4 $\mu m$ using low-resolution spectroscopy \citep{mebrown1998,mebrown2000}. Water-ice absorption features were detected and interpreted to suggest similarity to Oberon and Umbriel, two large Uranian satellites. Such a link would argue for Nereid to have formed as a regular satellite which was scattered by Triton’s capture instead of a capture from a heliocentric orbit. However, the presence of water ice is not a conclusive basis to link outer solar system populations, as water ice has since been detected across a variety of TNOs and Centaurs \citep{barucci2011,mebrown2012}.

An independent investigation by \citet{rhbrown1999} also found water ice to be present in the 1.4-2.5 $\mu m$ spectrum of Nereid, but measured a blue-sloped continuum at a lower signal-to-noise ratio (SNR) than the \citet{mebrown1998} data. \citet{rhbrown1999} found the spectrum with the blue continuum to be well described by models of water ice mixed with a featureless and unidentified blue-sloped darkening agent. They interpreted this model to suggest an inherent difference between the opaque material on Nereid and the opaque material that creates the typically reddish colors of TNOs. However, their measurements were made using a C-type asteroid in place of solar analog star to measure reflectance, which could lead to spurious spectral slopes. The neutral slope measured by \citet{mebrown1998} would not have the same implication, as a water-rich spectrum with a neutral continuum slope implies some degree of reddening to suppress the blue-sloped continuum of water ice. 

While understanding the water absorption band strengths allows for some degree of comparison with other small bodies in the outer solar system, determination of the spectral continuum is also required to assess spectral similarities between Nereid and any other population. This work provides independent measurements of the reflectance spectrum of Nereid over a wider wavelength range of 0.8-2.4 $\mu m$. In doing so, we provide a detailed measurement of the spectral continuum in context with both detected ice absorption bands. To provide a framework for comparisons with other small bodies, we model the volatile and opaque components of Nereid’s surface using intimate mixtures following the formalism of \citet{hapke_2012}. While we do not attempt to uniquely determine non-volatile components, we explore a variety of spectral models to assess the restrictions that can be placed on Nereid’s surface properties.

\section{Methods} \label{sec:methods}
\subsection{Data Collection}
Observations were conducted over two nights (Sept. 30 and Oct. 29, 2019 UTC) using the SpeX instrument on the NASA IRTF \citep{rayner2003}. On both nights, Nereid’s visual magnitude was 18.7, with phase angles of 0.7 and 1.4 degrees, respectively. Spectra were collected in low-resolution prism mode with a 0.8 arcsecond wide slit, providing a spectral resolution of R~100. Our processing and reduction methodology, including correction for time-variable telluric features, makes use of the Spextool pipeline \citep{cushing2004} and is outlined in \citet{sanchez2013} and \citet{sanchez2015}. Across both nights, the summed flux represents 4 hours and 40 minutes of total integration time. We observed a local G0V standard star once every ~80 minutes to perform telluric corrections. On both nights, we observed SAO 146541 for this purpose. We corrected the flux to solar colors by observing SAO 93936 on both nights. Seeing varied from 0.4-0.8 arcseconds, and relative humidity from 10-30\%.

Two main sources of systematic uncertainties can impact our analysis and modeling work. The first is related to wavelength-dependent slit losses, which can affect the continuum slope in our spectra. Due to the long integration times required to observe Nereid (which is near the faintness limit for our observing mode), the per-frame SNR is quite low. Therefore, assessing whether there is a systematic variation in the spectral slope requires collection of many individual frames to check for spectral variation with as a function of airmass or slit alignment. Nereid’s slow on-sky motion helps mitigate issues from tracking errors, but such errors cannot be fully eliminated. All frames were checked to measure their flux at K band, with poor frames identified if K band fluxes were consistent with zero. During the first night of observations, Sept. 30, 48 frames (all of 200s exposure time) were collected, and 6 were discarded. On Oct. 29, 48 additional frames were collected, and six frames were discarded as well. Slit alignment was kept to within $\sim$20 degrees of the object’s parallactic angle at all times. Any slit misalignment would largely manifest by changing the slope at z and J bands as a function of time/airmass. Within uncertainty, no trend in spectral slope was observed with airmass or slit alignment.

\begin{figure}[ht!]
\plotone{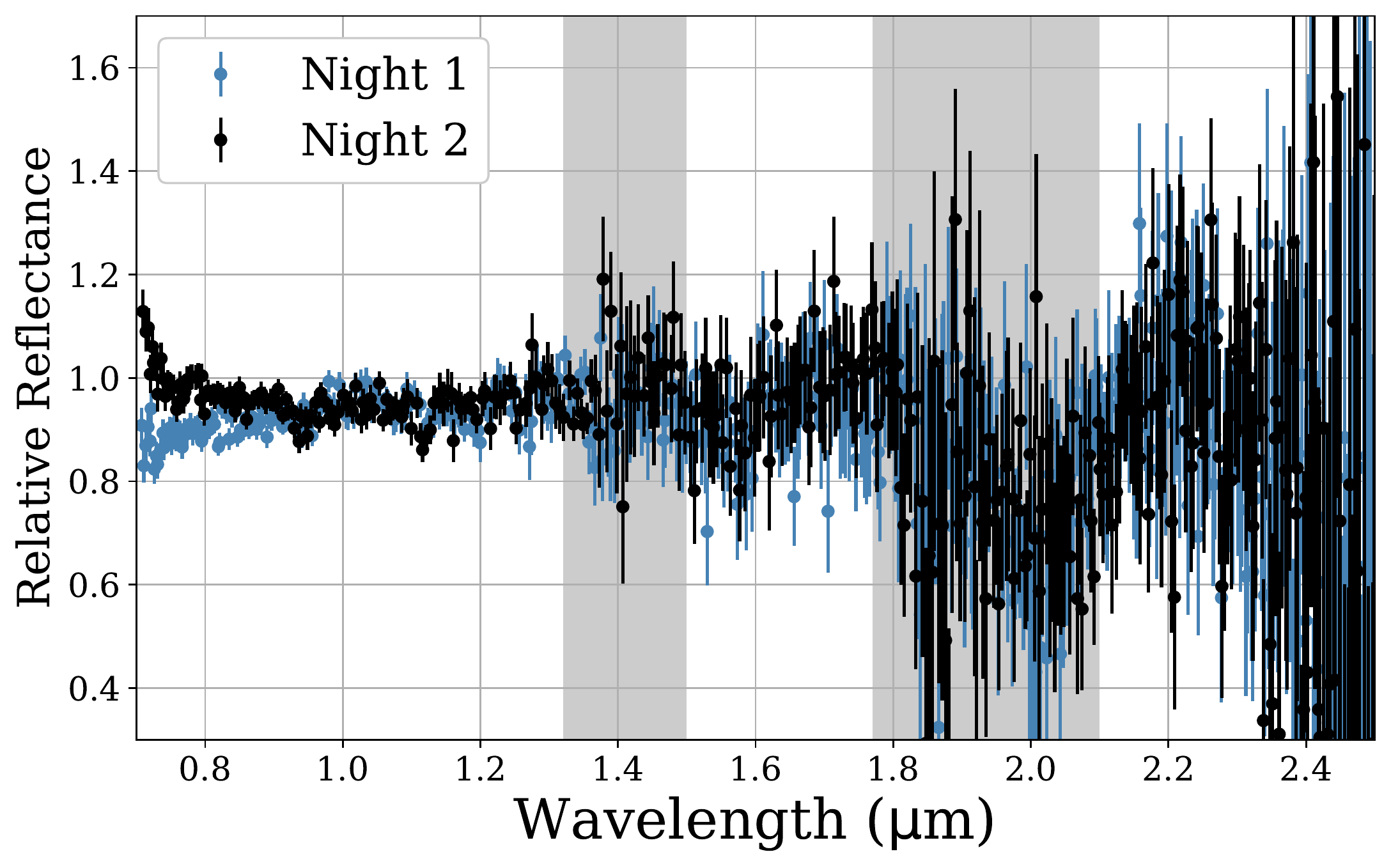}
\caption{Comparison of averaged reflectance spectra from the two nights, each normalized to unity at 2.2 $\mu m$. The presence and structure of both water ice absorption bands is consistent across both nights. The two spectra have close point-by-point agreement (within one-sigma, unbinned) throughout the full wavelength range. Divergence is noted at wavelengths below 0.85 $\mu m$, which are likely due to instrumental effects from the dichroic cutoff at 0.7 $\mu m$. Critically, the depths of both water absorption bands (1.5- and 2.0-$\mu m$) are in strong agreement, as is the overall slope from 1.2-2.2 $\mu m$, which are the main spectral features considered in our modeling. Data shown is binned by a factor of two for clarity.}
\label{fig:rel_comp}
\end{figure}

A second source of systematic uncertainties derives from the removal of telluric absorption features. This is of particular concern due to the overlap in wavelength between prominent telluric water vapor absorption features and the 2.0-$\mu m$ absorption band of solid water ice. In order to determine if the telluric removal procedure is robust, we intentionally split our observations across two different nights. Figure \ref{fig:rel_comp} displays the summed spectrum from each night, with the wavelength regions of strong telluric absorption features indicated with gray shading. We note strong agreement between observed band depths on both nights. We find agreement in the spectral slope measured by both data sets within the 4.2\%/$\mu m$ threshold for systematic effects determined by \citet{marsset2020}. We also note close agreement between our observed 2.0-$\mu m$ band depth and that measured by \citet{mebrown2000}. 

With the self-consistency of the two nights of observations confirmed, modeling and analysis was performed by combining all data into a single, averaged spectrum. This reflectance spectrum is collected in relative flux. We convert relative flux to geometric albedo by scaling to the measured 0.72 $\mu m$ albedo of $0.18^{+0.08}_{-0.05}$ as measured by \citet{buratti1997}, whose uncertainty is dominated by the size measurements of Nereid from Voyager observations \citep{thomas1991}.

\subsection{Modeling Approach}

Our goal is to provide what we regard as the simplest mixing model which well describes all major observed spectral features. In this data set, that means fitting band shapes, centers, depths, and the continuum. To determine what physical constraints our data can place on Nereid’s surface properties, we utilize the Markov-chain Monte Carlo (MCMC) Hapke modeling methodology after \citet{sharkey2019}, which utilizes the emcee python package \citep{foremanmackey2013} and makes use of methodologies developed by \citet{emery2004} and \citet{emery2011} for interpreting models of featureless asteroid spectra. Our aim is to explore the types of materials which constitute plausible analogs to a given surface. Besides absorption features clearly attributable to water ice, no other diagnostic features were observed within the SNR of our observations. This means that our choice of darkening components, when combined with water ice, must match Nereid's measured spectral continuum and its visual albedo, without suppressing the observed absorption features or introducing features other than those of water ice.

Generally, our modeling in this work is insensitive to moderate variations in visual albedo. Variations in the albedo of 0.10 or more from its nominal value can be accounted for by changes in the mixing ratio of water ice and a given darkening agent. This is allowed due to the grain size dependence of water ice absorption band depth (increasing depth with larger grain size). We note that even for a fixed albedo, deriving absolute abundances of materials in these mixtures is highly model dependent, and should not be regarded strictly. Instead, abundances are useful as far as they provide a benchmark for Nereid’s composition that can be readily compared with other objects observed at similar wavelengths, and to identify the range of spectral diversity which can be explained by a single set of compositional endmembers.

A series of models, each of varying complexity, were performed to determine the minimum number of Hapke parameters required to describe Nereid’s spectrum. Figure \ref{fig:ice_comp} shows the data binned by a factor of 15. Two-component models, such as those given in Figure \ref{fig:ice_comp}, describe the surface as a simple combination of an opaque material combined with either amorphous or crystalline water ice. This binned spectrum is useful to illustrate the overall SNR of the different features we detect. Since the choice to bin implicitly makes an assumption about the width of any features in the data set, all models were performed on the unbinned spectrum. In all models, a single grain size for all materials is assumed. As with \citet{sharkey2019}, we find that allowing for unique grain sizes for each component produces model degeneracies and slows convergence without improving fits. We adopt a single grain size bounded over the range of 0.5-1000 $\mu m$.

Three component mixing models that include at least one opaque component and at least one form of water ice provide stronger fits than the two-component models. Attempts to add additional components did not improve fits to the water ice absorption bands. Since the absorption bands are the only component of the spectrum that are not model dependent, we conclude that adding additional trace mixing components are not justified by the NIR data alone. This includes models with multiple types of water ice with differing grain sizes, as well as models with many different darkening agents. Optical constants for 60K amorphous and crystalline water ice are taken from \citet{mastrapa2008} and used in our models over a wavelength range of 1.2-2.5 $\mu m$. Optical constants from \citet{warren1984} were used in our models across 0.7-1.2 $\mu m$. Models which used optical constants of water ice with differing temperatures from 40-80K were not found to impact fit quality, as small grain size or continuum adjustments can be made to cancel out absorption band variations. Optical constants for amorphous carbon are taken from \citet{rouleau1991}, Tagish Lake from \citet{roush2003}, and magnetite from the JENA database.\footnote{https://www.astro.uni-jena.de/Laboratory/OCDB/oxsul.html}

\section{Results} \label{sec:results}

As can be seen in Figure \ref{fig:rel_comp}, we measure the NIR spectrum of Nereid to have a slightly reddish slope, with its reflectance increasing by ~10\% from 0.8-2.2 $\mu m$. We detect the two strongest water ice absorption band complexes, centered near 1.5 $\mu m$ and 2.0 $\mu m$, across both nights at consistent band depths. We also detect a 1.65-$\mu m$ absorption band on both nights, which is diagnostic of crystalline water ice.  Overall, we measure the spectrum of Nereid to be in good agreement with the full range of the 1.4-2.4 $\mu m$ spectrum reported by \citet{mebrown2000}, including the detection of the 1.65-$\mu m$ band due to water ice.  With our measurement of Nereid's reflectance continuum from 0.8-2.4 $\mu m$ to be increasing (red-sloped), we will use spectral models to explore several different darkening/reddening agents.

For ease of reference, and to assess the significance of our band detections and subsequent fits, we provide depths and centers of these three bands in Table \ref{tab:bandpars}. Due to the low per-pixel SNR of our spectrum, we calculate observed band properties by first smoothing the spectrum with a moving-average filter. The band depth is determined as the minimum weighted-average value within the smoothed band, relative to the reflectance value at 2.2 $\mu m$ (the wavelength at which our spectrum is normalized). This calculation returns band depths of $13\pm2\%$ for the broad 1.5-$\mu m$ water ice band, $10\pm3\%$ for the 1.65-$\mu m$ band diagnostic of crystalline water ice, and $27\pm3\%$ for the 2.0-$\mu m$ water ice band. The band minimum is determined as the location of this minimum reflectance, with its uncertainty given as the wavelength range that is within one standard deviation of the minimum reflectance value. The band minima are constrained in this way to $1.5_{-0.02}^{+0.02} \mu m$, $1.65_{-0.01}^{+0.02} \mu m$ , and $2.01_{-0.02}^{+0.01} \mu m$.While this does not include any continuum removal (and therefore does not provide formal band parameters), it does give clear descriptions of our spectrum which can be compared with our spectral models. 

\begin{figure}[ht!]
\plotone{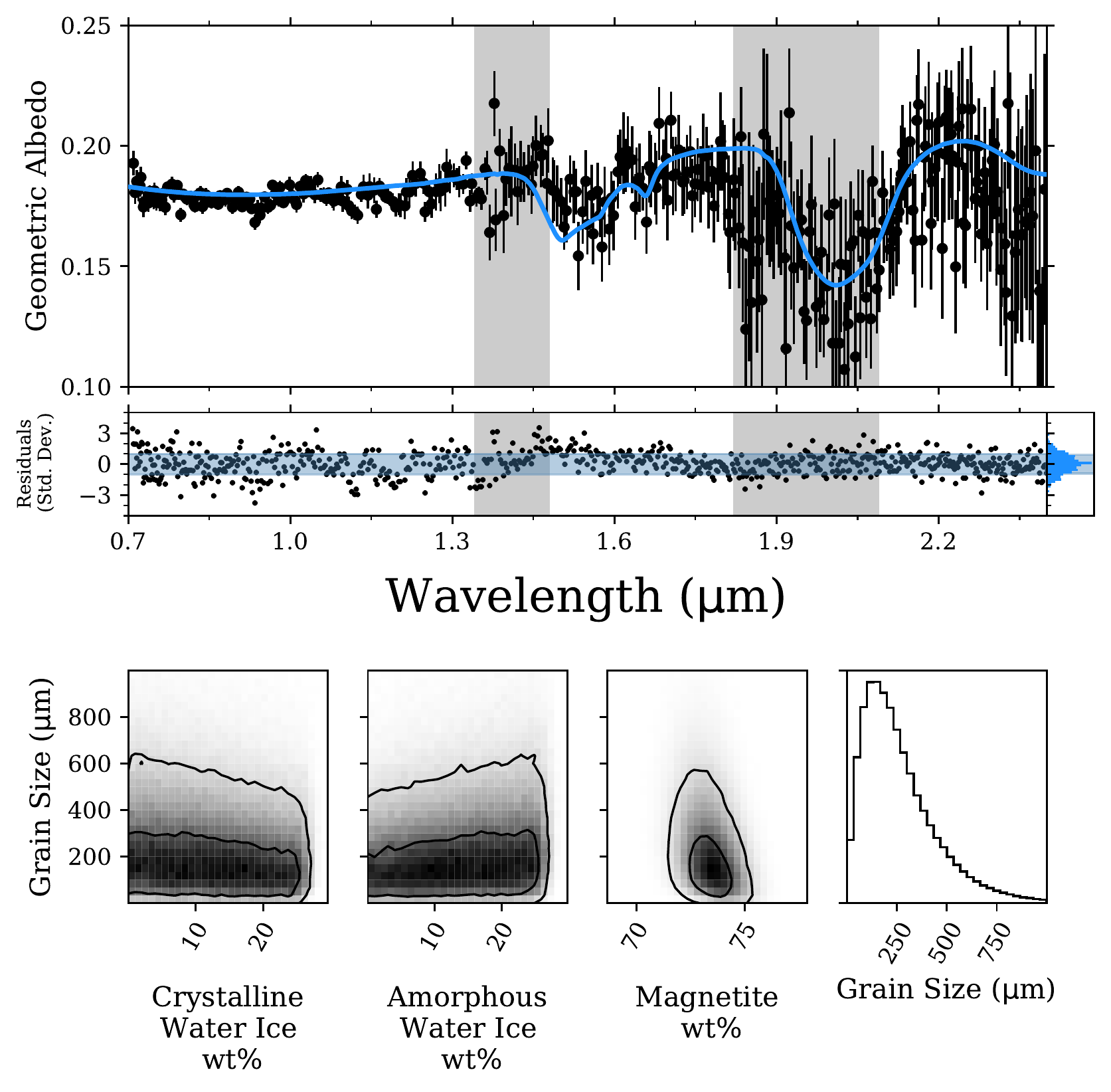}
\caption{Best fit models for mixtures of water ice with magnetite as the darkening agent. This shows that the spectral continuum of magnetite provides a good fit to the observed data, including good simultaneous fits to both the 1.5-$\mu m$ and 2.0-$\mu m$ water ice absorption bands. We note that the continuum from 0.8-1.3 $\mu m$ does not in and of itself discriminate between darkening agents. Instead, this conclusion is only apparent by evaluating fits over the full wavelength range. Returned posterior mixing distributions are shown in the bottom panels. While the mixing ratios of ice vs. opaque materials is well constrained regardless of grain size (note that compact distribution of the mixing percentage of magnetite), no strong preference is found between crystalline or amorphous ice when fitting both the 1.5-$\mu m$ and 2.0-$\mu m$ bands simultaneously. Note that the distribution of ice mixing ratios includes models that are purely crystalline and purely amorphous. }
\label{fig:H2OWMag}
\end{figure}

\begin{figure}[ht!]
\plotone{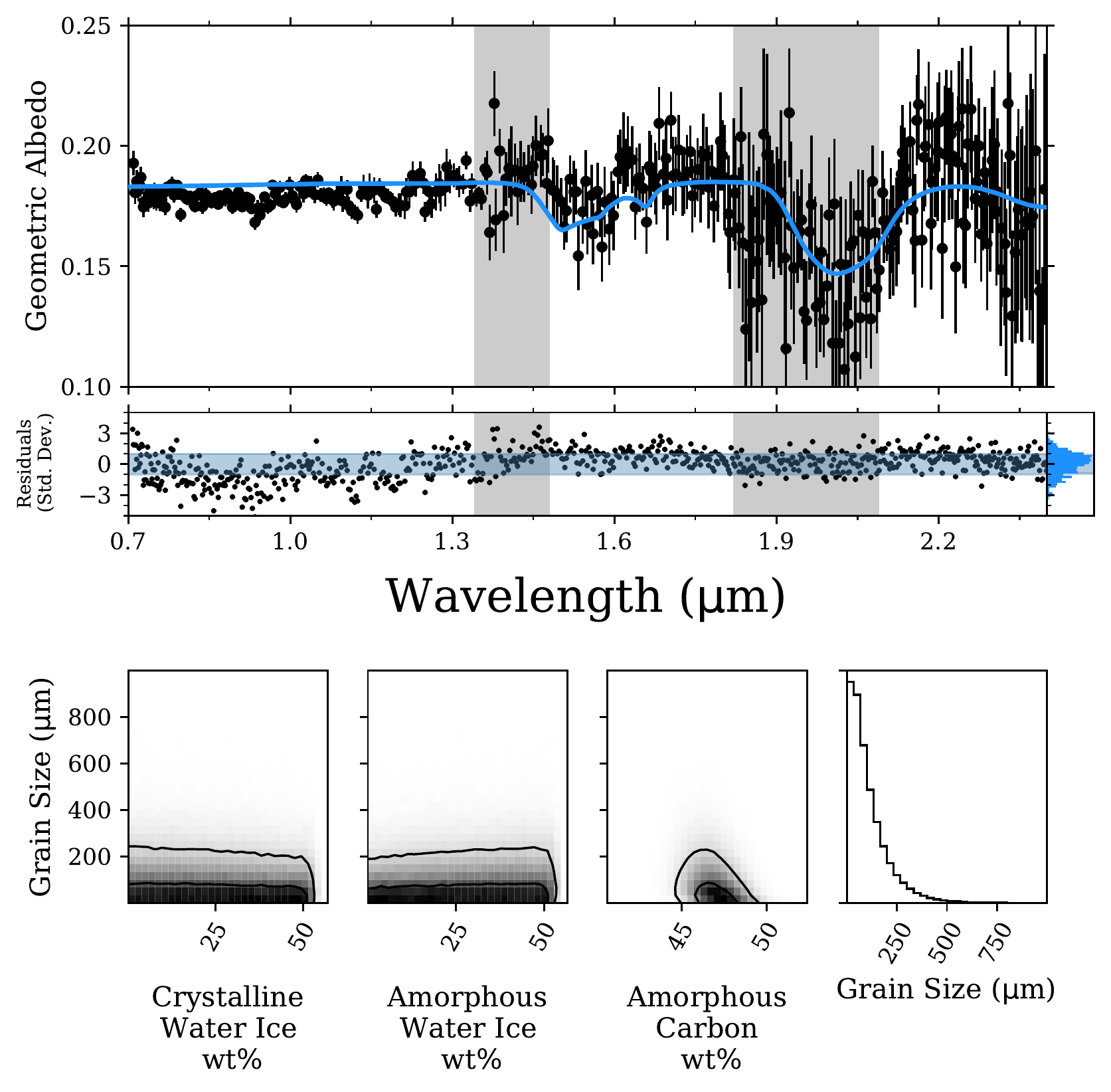}
\caption{Best fit models for mixtures of water ice with amorphous carbon. The neutral-sloped continuum provided by amorphous carbon as the lone darkening agent gives a poor fit to the two water band depths or shapes. The misfit is most apparent in the shoulder near 2.2 $\mu m$. Returned posterior mixing distributions are shown in the bottom panels. Again, the mixing ratios of ice vs. opaque materials is well constrained, and grain sizes are constrained towards the lowest end of our modeled range (note that narrow distribution of amorphous carbon mixing percentage and the narrow distribution of returned grain sizes). As in Fig. \ref{fig:H2OWMag}, no strong preference is found between crystalline or amorphous ice when fitting both the 1.5-$\mu m$ and 2.0-$\mu m$ bands simultaneously.The distribution of ice mixing ratios includes models that are purely crystalline and purely amorphous.}
\label{fig:H2OWD}
\end{figure}

\begin{figure}[ht!]
\plotone{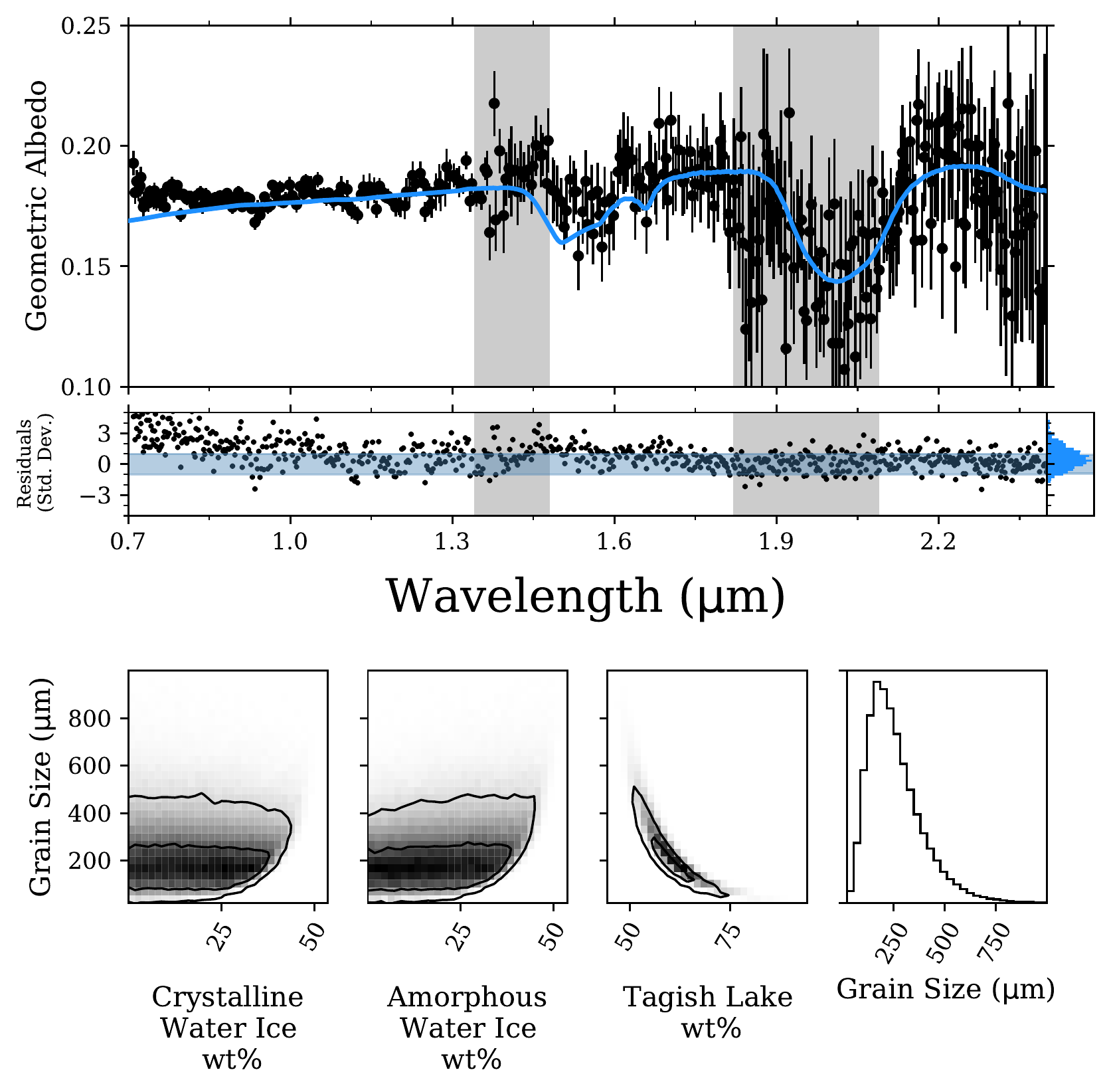}
\caption{Best fit model of water ice mixed with Tagish Lake as the darkening agent. This model shows weaker fits to both the 1.5 $\mu m$ band, as well as the shoulder near 2.2 $\mu m$. This illustrates that a spectral continuum that is approximately linear is disfavored by our data. Returned posterior mixing distributions are shown in the bottom panels. The mixing ratios of ice vs. opaque materials is constrained but highly dependent on model grain size (note the correlation between higher mixing ratios of Tagish Lake and lower grain sizes). As in Figs. \ref{fig:H2OWMag} and \ref{fig:H2OWD}, no strong preference is found between crystalline or amorphous ice when fitting both the 1.5-$\mu m$ and 2.0-$\mu m$ bands simultaneously.The distribution of ice mixing ratios includes models that are purely crystalline and purely amorphous.}
\label{fig:H2OWTag}
\end{figure}

To interpret how strongly our observations can constrain the darkening material(s) present on Nereid, we ran several different mixing models using dark opaques with different spectral properties. Figures \ref{fig:H2OWMag}-\ref{fig:H2OWTag} display model fits of three component Hapke mixing models that incorporate various combinations of amorphous and crystalline water ice (measured at 60K), plus one darkening agent. Each figure displays the spectrum, binned by a factor of two, compared with the best-fit model in the top panel, with point-by-point residuals displayed in the middle panel and the residual distribution displayed alongside as a histogram to aid comparisons.

\begin{figure}[ht!]
\plotone{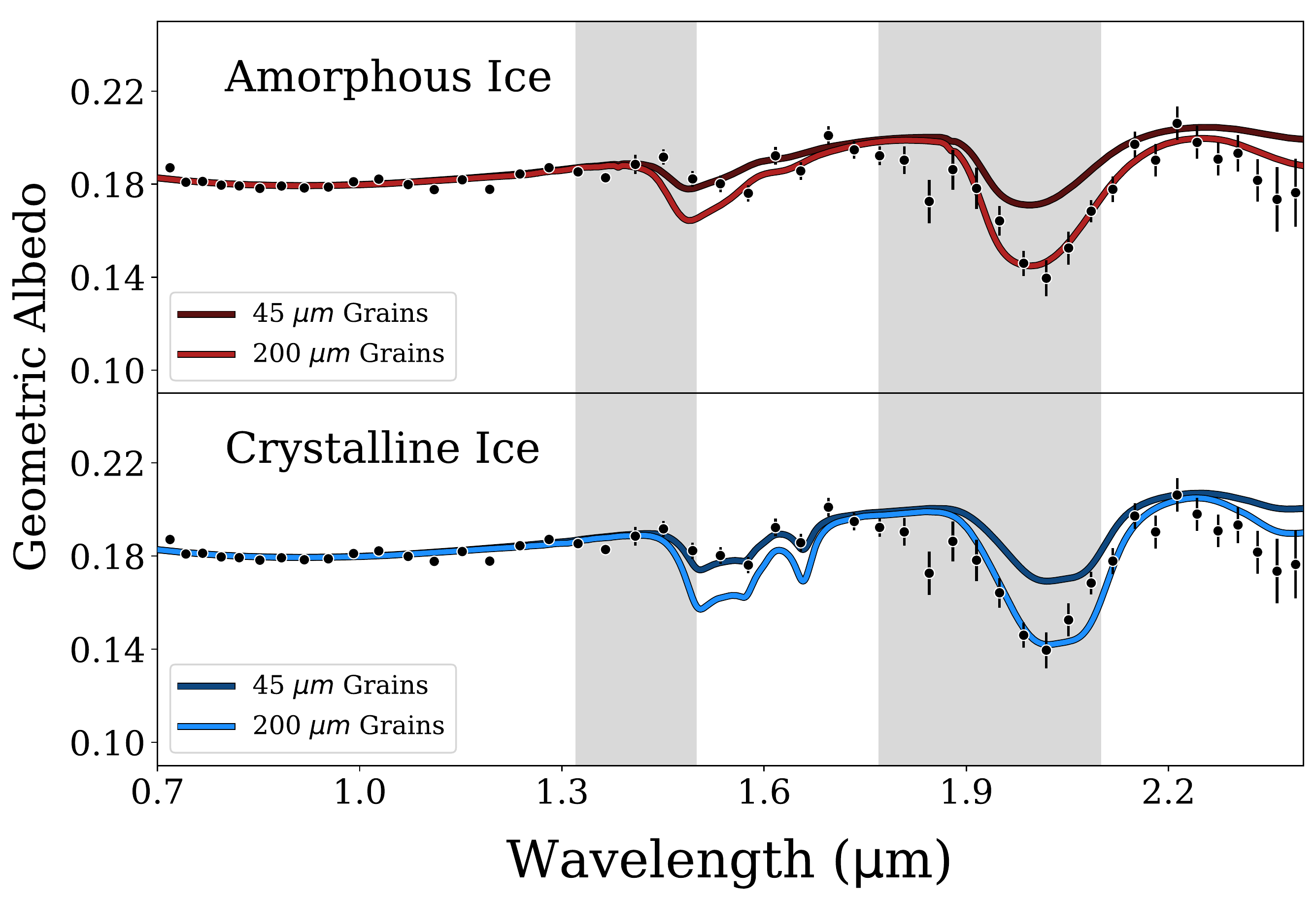}
\caption{Comparison of end-member water ice models mixed with magnetite. A single model with crystalline water ice (lower panel) can be tuned via grain size adjustments to provide good fits to the depths of either the 1.5-$\mu m$ or 2.0-$\mu m$ absorption features, but not simultaneously. Notably, the shape and center of the 2.0-$\mu m$ feature is better fit by models using amorphous ice (upper panel). Behavior of our mixing models are largely a function of the tradeoff between fitting either band, resulting in best fit models that prefer ice mixtures instead of a single phase. Measurements of the 2.0-$\mu m$ band are complicated by the presence of significant telluric absorption over this region. Data is shown binned by a factor of 15 to demonstrate the significance of detected absorption features. Both models use mixing ratios of water/magnetite of 0.265/0.735, and use grain sizes of either 45 $\mu m$ or 200 $\mu m$ as indicated.}
\label{fig:ice_comp}
\end{figure}

Models are presented with the posterior distributions of mixing parameters which were allowed to vary. The mixing ratios of the three-component models are shown as two-dimensional distributions vs. grain size. The shading of the two-dimensional distributions corresponds to the number of model steps spent within a single bin, with contours showing the one- and two-sigma uncertainty ranges. The total distribution of model grain sizes is displayed as a one-dimensional histogram. Critically, the posterior distributions show that the mixing models return smooth, single-peaked solutions. The posterior distributions can also be used to show the tradeoffs that can be made between mixing ratios of various components as a function of grain size as well as the flexibility of each model. For example, while the opaque component in Figs. \ref{fig:H2OWMag}-\ref{fig:H2OWD} are constrained to a single best-fit value with little grain size dependance, Fig. \ref{fig:H2OWTag} shows that the mixing ratio of Tagish Lake can vary from $\sim 50-75\%$ with considerable dependence on grain size. Additionally, the mixing ratio distributions in \ref{fig:H2OWMag}-\ref{fig:H2OWTag} show that, when models include both crystalline and amorphous ice, no strong preference between either is returned. This lack of a preference shows the effect of the tradeoff between fitting the $1.5 \mu m$ and $2.0 \mu m$ absorption bands as discussed and shown in Figure \ref{fig:ice_comp}.

We find that Hapke fits of the relative position and depth of the water ice bands are improved by including both amorphous and crystalline water ice (compare individual fits in Fig. \ref{fig:ice_comp} with the fits displayed in Figs. \ref{fig:H2OWMag}-\ref{fig:H2OWTag}). While this finding may imply novel properties of the water ice present on Nereid's surface, from the perspective of assessing the darkening material, all that is relevant is that this combination of ices improves fit quality and provides flexibility for the models to explore different band properties.

For models which use only a single darkening agent, the use of magnetite was found to provide the best single spectral match to the telescopic data. While some magnetites have a broad local minimum in reflectance near 1.0 $\mu m$ \citep[see discussion in][]{izawa2019}, we note that the reason magnetite is preferred is not based on any curvature detected from 0.7-1.3 $\mu m$. As can be seen in the fits for magnetite (Fig. \ref{fig:H2OWMag}) vs. the fits for amorphous carbon (Fig. \ref{fig:H2OWD}), both components can provide good quality fits across the 0.8-1.3 $\mu m$ range. Instead, the preference for magnetite is due to its better match to the overall reflectance from 1.3 to 2.2 $\mu m$. This improved match to the spectral continuum also results in improved fits to the 1.5-$\mu m$ ice absorption band, whose depths are overestimated when using more neutrally colored materials to fit the stronger 2.0-$\mu m$ band (Figs. \ref{fig:H2OWD} and \ref{fig:H2OWTag}).

\begin{figure}[ht!]
\plotone{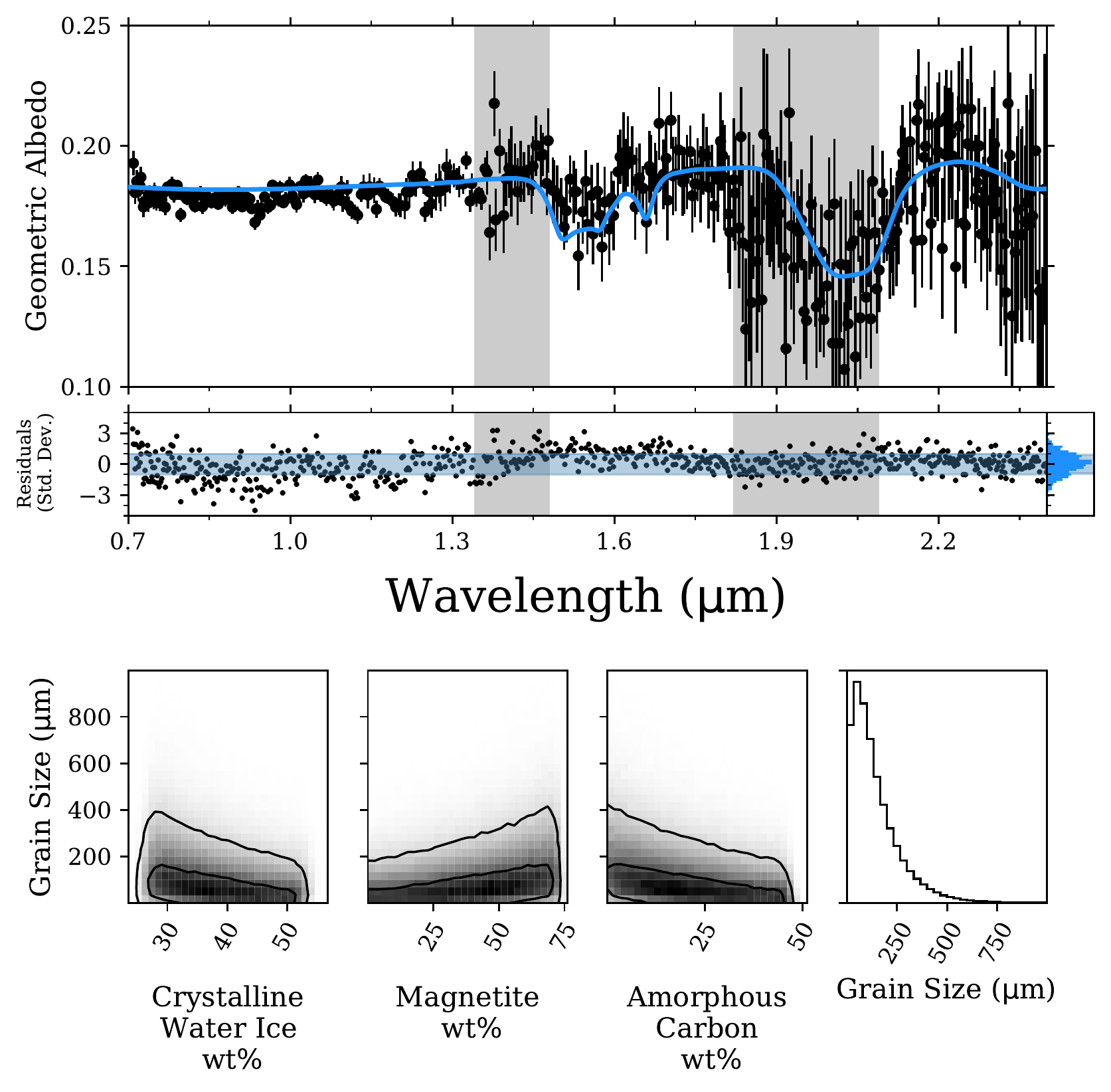}
\caption{Comparison of a purely crystalline water ice model mixed with magnetite and amorphous carbon as mild reddening and darkening agents. Under these end-member assumptions, the model provides a good match to the continuum slope as well as the band depth and positioning of the 2.0 $\mu m$ feature. However, this model overestimates the band depth of absorption bands at 1.5 $\mu m$. Such a tradeoff is typical of our models. Behavior of our mixing models are largely a functional tradeoff between fitting either band, which results in best fit models that prefer ice mixtures instead of a single phase. Measurements of the 2.0-$\mu m$ band are complicated by the presence of significant telluric absorption over this region. Returned posterior mixing distributions are shown in the bottom panels. Here, the mixing ratios of ice vs. opaque materials is considerably less constrained than those of Figs. \ref{fig:H2OWMag}-\ref{fig:H2OWTag} (note the wide distribution of crystalline water ice mixing percentages). Grain sizes are constrained towards the lowest end of our modeled range.}
\label{fig:H2OMagD}
\end{figure}

In principle it is possible to construct an arbitrarily complex continuum in an attempt to perfectly fit both water ice absorption bands. However, such a fit is not guaranteed to be found, nor is it guaranteed to be unique if it is found. In particular, this is a problem in the absence of having detailed knowledge of the water-ice features (requiring higher SNR), as these absorption bands are the only diagnostic features in the data. However, the question of the choice of continuum materials is essential to understand whether or not a complex mixture featuring both amorphous and crystalline ice is uniquely supported by our observations or not. Figure \ref{fig:H2OMagD} illustrates a continuum which allows for crystalline ice alone to provide a closer match to the observed 2.0-$\mu m$ band. Although the band depth and position of the 1.5-$\mu m$ feature is in good agreement with the data, the depth and position of the 2.0-$\mu m$ band are underestimated. Therefore, regardless of the assumed form of water ice, we show that both absorption bands cannot be simultaneously explained by a single form of water ice in our models.

Given that our purpose is to characterize the ice absorption features in tandem with the overall spectral continuum, we deem the fit quality of the explored models to be sufficient. Notably, the magnetite model which combines both crystalline and amorphous ice returns the lowest $\chi_R^2$ value of any of our tested runs (Table \ref{tab:model_comps}). This model gives $\chi_R^2=1.21$, compared with $\chi_R^2=1.36$ for crystalline water ice with carbon and magnetite. Other models with both amorphous and crystalline ice perform worse: the neutral continuum of amorphous carbon results in  $\chi_R^2=1.67$, and the linearly red-sloped continuum of Tagish Lake provides $\chi_R^2=1.55$. We stress that the exact choice of darkening agent is arbitrary. However, it is noteworthy that magnetite alone provides a strong spectral continuum match. We also note that the use of magnetite as an endmember illustrates that, while some type of reddening material is required to reproduce Nereid's neutral-sloped spectrum (as water-ice is blue-sloped), Tholin-like organics are not required. Assessing model behavior using exact $\chi_R^2$ values alone can be complicated by the decreased SNR at longer wavelengths (causing uneven weighting of the data), but using this metric along with the other parameters in Table \ref{tab:model_comps} provides a comprehensive summary of the impact of changing mixing components.

As noted previously, model behavior is generally insensitive to the precise albedo assumed. This is because the albedo is largely a product of the mixing ratio of ice vs. darkening agent, while the ice absorption band depths for a given mixture can be tuned by adjusting the grain size. Models were performed that allowed for different grain sizes of each component, i.e. different textures for the ices than for the darkening agent. These models do not produce improved fits, but only increase computation times and provide more complex posterior parameter distributions. Fits do not improve if we include albedo as a free parameter that can vary by 0.10. 

By comparing the fit quality of our various end-member models, several general conclusions can be formed (refer to summary of fit qualities in Table \ref{tab:model_comps}). Models can easily fit one of the two water ice absorption bands but cannot provide perfect fits to both simultaneously using our mixing parameters. This may be a product of more complex mixing scenarios than the intimate mixtures considered here or be related to properties of water ice on Nereid which are distinct from laboratory materials.

Simply mixing water ice with a spectrally neutral darkening material does not provide optimal fits to Nereid's slope. This point shows that the water-ice content required to match the absorption features is high enough to affect the slope of any of our models. In other words, water ice bands cannot be simply overlaid upon a neutral continuum. This property is not derived from the relative reflectance spectrum on its own but is only apparent by taking into account the albedo value of 0.18.

The relative strengths and positions of the measured water ice absorption features are difficult to describe with a single spectral model. Specifically, the 1.5-$\mu m$ features are consistently weaker than would be expected from our simulated Hapke mixtures. This issue can be mediated by including amorphous ice into our models, as this effectively "spreads out" the 1.5-$\mu m$ features without similarly weakening the 2.0-$\mu m$ feature. Such a solution is very useful for our models but is not unique. Determining whether or not such a solution is truly warranted will require additional modeling efforts which we do not attempt - our models are meant only to illustrate the existence of this problem.

Regardless of adjustments to the relative strengths of the 1.5 and 2.0 $\mu m$ bands caused by micron and sub-microns sized grains, our models show evidence that additional modification to the shape of the absorption bands is required (compare the quality of the two-component models to observed absorption bands in Figure \ref{fig:ice_comp}). This behavior persists in the three-component models, as fits do not return a preference between crystalline and amorphous ice across any continuum fit explored in Figs. \ref{fig:H2OWMag}-\ref{fig:H2OWTag}. Since crystalline ice alone provides adequate fits to the 1.5 $\mu m$ band, we conclude that the 2.0 $\mu m$ band is the main driver of this behavior. While fitting a continuum with both amorphous carbon and magnetite improves models that use only crystalline ice (Fig. \ref{fig:H2OMagD}), the discrepancy in 2.0 $\mu m$ band shape persists (compare the derived 2.0 $\mu m $ band minima shown in Table \ref{tab:bandpars}).

\section{Interpretations} \label{sec:interp}

Our most significant finding is that it is challenging to fit the spectrum of Nereid using a single type of water ice. This discrepancy is only revealed by measuring the spectral continuum at shorter wavelengths to fully constrain the continuum slope, and therefore provide better context for measured band properties. Several scattering scenarios exist which could satisfy our measurements. The first is that the continuum is adequately described by a single darkening agent such as magnetite, which when combined with water ice features a reddish slope from 1.0-1.7 $\mu m$ and a more neutral slope from 1.7-2.2 $\mu m$. In this case, we are unable to fit a single type of water ice (amorphous or crystalline) to both absorption bands with an intimate mixture based on Hapke formalism. This scenario allows for more complex mixtures. One example may be finer sized grains than we assume. \citet{yang2014} report that when using a Mie-scattering reflectance model, the ratio of 1.5- and 2.0-$\mu m$ band depths becomes a grain-size dependent parameter. This solution would be consistent with findings from studies of icy satellites in the Saturnian \citep[e.g.,][]{Clark2012,Scipioni2017} and Uranian \citep[e.g.,][]{cartwright2018,Cartwright2020} systems which incorporated sub-micron/micron scale particles (along with larger grains) to more accurately model these H2O ice bands and other spectral features. However, simply adjusting the relative band depths of our crystalline ice model do not reproduce the measured shape of Nereid's 2.0 $\mu m$ band (as displayed in Figure \ref{fig:ice_comp}). Therefore, we conclude that an alternative or additional effect is supported by our observations.

Another valid interpretation is to adjust the spectral continuum to be neutral from 1.5-2.4 $\mu m$. This would allow for fits of crystalline water ice to better match the 2.0 $\mu m$ band, but would decrease the fit quality outside of the band. This scenario still cannot explain the reduced depths of the water ice absorption features, as seen in Fig. \ref{fig:H2OWD} and Table \ref{tab:model_comps}. Again, a more complex mixing scenario would be necessary in order to describe this spectrum. 

In order to provide a satisfactory fit with crystalline ice alone to the measured absorption bands, a blue continuum must be invoked. This would imply that the spectrum measured from 0.8-1.3 $\mu m$ does not represent a featureless continuum, but instead indicates the presence of material(s) with broad absorption features in this region. This scenario requires the presence of additional unknown components: even magnetite, which exhibits such spectral behavior in this region, is found to give a reddish slope which does not satisfy this scenario.

We note that another interpretation could be due to a distribution of ice crystallinity. The 2.0-$\mu m$ band is very well described by models which incorporate amorphous ice as opposed to crystalline ice. Such a scenario is difficult to immediately rule out, as the expected surface temperature of Nereid (equilibrium temperature ~50-60K) is within the temperature regime where \citet{mastrapa2006} find that radiolytic production of amorphous ice retains some spectral properties of crystalline ice. With their detection of crystalline ice on (50000) Quaoar, \citet{jewitt2004} noted that the survival timescale of crystalline ice within the observable skin depth of a surface is $\sim 10^7$ years at the orbit of Neptune. Therefore, it cannot be assumed that only a single type of ice would be able to survive on the surface of Nereid. This scenario could manifest as an intimate mixture of both amorphous and crystalline ice, or as discrete surface units, perhaps relating to differing exposure ages.

In the absence of further modeling, it is useful to provide an initial comparison between Nereid and both icy satellites and small bodies whose orbits interact with Neptune. These comparisons, in spanning a large size range and hence geological conditions (i.e. whether differentiation occurred), provide an assessment of the spectral variety which must be explained in order to associate Nereid with any other groups of outer solar system objects. Further spectroscopic observations of such groups would provide context for interpreting the results of broadband color surveys which have shown critical links between various small body populations. 

Nereid lacks the red sloped continuum typically seen on large TNOs such as Pholus \citep{cruikshank1998} and Quaoar \citep{jewitt2004}, but it is not totally anomalous. In particular, the spectrum of (90482) Orcus has a neutral slope similar to Nereid that was found to be well-described by  models that utilize blends of amorphous and crystalline ice \citep{DeMeo2010}. However, such models required incorporation of an additional blue component, so comparisons to our modeling methods are not direct. In broad comparison with KBOs and Centaurs, Nereid’s spectrum falls within the neutral (least-red) tail of the KBO color distribution \citep{marsset2019}. The spectrum of (136108) Haumea, a TNO with an albedo of $0.51\pm0.02$ and an ellipsoidal shape of $\sim1700\times1100$ km \citep{ortiz2017}, was found to best match a pure crystalline water ice spectrum, but also requires an additional blue component \citep{trujillo2007}. Nereid’s reflectance properties, particularly its spectral slope, do not directly match one single analog among large TNOs, but the spectral variety of this group makes it difficult to assess Nereid’s affinity.

Amongst small satellites in the outer solar system, several objects stand out with suitably dark ($\sim 20\%$ albedo) surfaces yet deep ($\sim 25\%$) water ice bands. The Saturnian satellite Hyperion has a steep red slope from 0.8-1.4 $\mu m$ attributed to the presence of organic material, but its surface includes low-albedo surface units with deep water-ice bands of comparable depth to Nereid \citep{Cruikshank2007}. The regular Uranian satellite Umbriel also has a similar albedo \citep[$\sim$0.21,][]{karkoschka1997}. Comparison between our spectrum for Nereid and the spectra of Umbriel collected by \citet{cartwright2018} is convenient as both observations were observed using the same instrument with similar processing techniques. These two objects have similar band depths, with Umbriel’s 1.5 $\mu m$ and 2.0-$\mu m$ band depths averaging 9\% and 26\%, respectively \citep[Table 7 of][]{cartwright2018}. The main difference is apparent in Umbriel’s blue-sloped spectrum from 1.4-2.5 $\mu m$. The similar band depths and albedo, but differing continuum properties from Nereid suggests differences in the darkening agent(s) present on these surfaces. However, Umbriel (diameter 1170 km) is significantly larger than Nereid (diameter 340 km), so whether this difference is related to intrinsic composition, surface textures, or different geological processing factors cannot be simply assessed. 

\begin{figure}[ht!]

\plotone{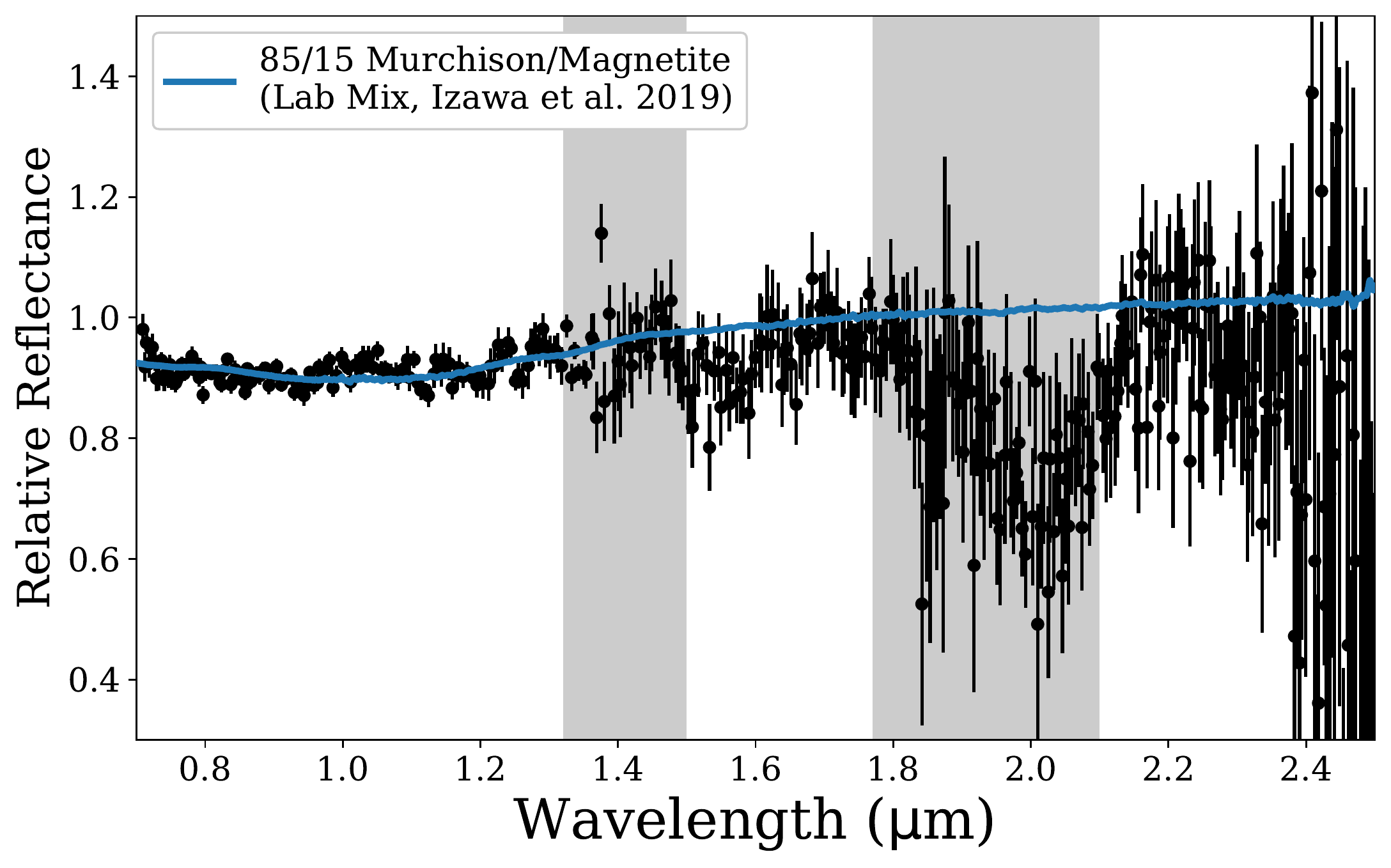}
\caption{An example of our spectrum compared with a continuum represented by a laboratory mixture of magnetite with Murchison. This shows that such materials provide an excellent baseline for interpreting neutral spectra. Note that in our mixing models, adding in water ice adds a blue component to the continuum, which is compensated in the fits by increasing the magnetite content and recreating a continuum that is similar to the laboratory mix shown here.}
\label{fig:labmix}
\end{figure}

The agreement between the spectrum of the non-ice components of Nereid’s surface and a mixture of magnetite and the CM2 carbonaceous chondrite Murchison is striking (Fig. \ref{fig:labmix}). While this spectral match is not diagnostic of a particular composition it is interesting to consider that CM chondrite material is widely distributed at least in the inner solar system, yet likely originated at much larger heliocentric distances. Analysis of stable isotopes, especially the chromium, titanium, and oxygen isotopic systematics, show that currently analyzed inner solar system materials accreted from a reservoir distinct from the carbonaceous chondrites, consistent with accretion of carbonaceous chondrites beyond the ice line \citep[e.g.,][]{warren2011} and subsequent redistribution during giant planet migration \citep[e.g.,][]{scott2018}. Magnetite in carbonaceous chondrites is almost entirely the product of aqueous alteration of primary mafic silicates (dominantly olivine and pyroxene). The presence of magnetite on the surface of Nereid could therefore imply the incorporation of some material that has undergone low-temperature aqueous alteration. Such material could have been incorporated during Nereid’s formation, or could be due to later contamination due to infall of material similar to aqueously altered carbonaceous chondrites, as has been observed on many planetary surfaces including Vesta \citep{Reddy2012,Nathues2014}, Iapetus \citep{Spencer2010,Tosi2010} and Earth's Moon \citep{Zolensky1997}. 

While our results should not be taken as evidence for the presence of carbonaceous-chondrite-like material on Nereid’s surface, they do suggest that carbonaceous chondrites and their components may be useful analogues for the non-ice components of spectrally neutral outer solar system bodies, as suggested recently by \citet{fraser2019}. Such analogues may provide a useful lens to interpret how color variation amongst small bodies in the outer solar system relates to compositional differences, and therefore speak to how well spectral measurements can constrain the materials that comprise these surfaces. 

One final caveat is that at present, there appear to be no reflectance spectral measurements for particulate magnetite below the Verwey transition temperature ($\sim$125K), which may cause subtle spectral differences. Spectral changes are plausible below this temperature because the crystal structure is known to change at the Verwey transition from cubic to monoclinic \citep[e.g.,][]{Wright2002}. Additionally, below the Verwey transition, magnetite changes from electrically conductive to electrically insulating due to the loss of electron delocalization \citep[e.g.,][]{Piekarz2007}. We expect any such changes to be subtle, but changes in crystal structure require acknowledging the potential for spectral changes which would have to be assessed via laboratory experiment.

\section{Conclusions} \label{sec:conclusions}

We report reflectance spectra of Nereid from 0.8-2.4 $\mu m$, detecting the two major sets of water ice absorption bands in the NIR (near 1.5 $\mu m$ and 2.0 $\mu m$) in good agreement with previous observations. We measure the continuum from 0.8-1.4 $\mu m$ for the first time, which we find to be necessary for detailed assessments of Nereid's surface. We demonstrate that the water-ice rich surface of Nereid can be modeled with simplified Hapke mixing models to constrain its surface composition. We find that its spectrum presents complexity beyond a simple assumption of a surface dominated by large grains of crystalline water ice mixed with a single opaque component. We present interpretations derived by comparing models that use a variety of possible materials, which are summarized below.

\begin{itemize}
\item Magnetite is found to be a suitable material in our model fits, as its red slope from 1.3-2.5 $\mu m$, when combined with water ice, provides a good match to the spectral continuum without requiring the presence of Tholin-like organics. We note in particular that a darkening agent comprised of magnetite combined with a neutral material, such as amorphous carbon or bulk samples of the CM2 chondrite Murchison, are useful for investigating the surfaces of low-albedo, ice-rich objects with similarly neutral colors to those of Nereid.
\item While the water ice absorption features near 1.5 $\mu m$ are well described by models of pure crystalline water ice mixed with a spectrally reddish continuum, the 2.0-$\mu m$ feature is difficult to simultaneously explain. Models of both crystalline and amorphous ice which fit the band depth of the 2.0-$\mu m$ feature generally over-predict the depth of the 1.5-$\mu m$ feature. Models using pure crystalline ice produce lower quality fits to the observed shape of the 2.0 $\mu m$ band, regardless of depth, when compared to models using pure amorphous ice.
\item Allowing for mixtures of both amorphous and crystalline ice may solve this discrepancy but would likely require achieving higher SNR observations (particularly of the 2.0 $\mu m$ band) and higher spectral resolution observations of the narrow 1.65 $\mu m$ band in order to separate out this effect from other scattering scenarios (such as the presence of sub-micron sized grains) or temperature-dependent changes to ice band properties which this work does not attempt to constrain. 
\item The presence of either crystalline or amorphous ice on the surface of Nereid could be understood from several different contexts, as different mechanisms could have acted on its surface over time to process one form to the other. Considering that Nereid is the third largest satellite in a chaotic planetary system, any constraints on its thermal history may imply complementary constraints on its impact or disruption history.
\item Our best model fits (lowest $\chi_R^2$) incorporate mixtures of both crystalline and amorphous phases of ice, due to the differences in the properties of the 2.0-$\mu m$ absorption feature. In the most general terms, this finding implies different properties between the laboratory ice samples used in our models and the ice which is present on Nereid’s surface. We further note that Nereid's surface temperature of $\sim60K$ is near a regime identified by laboratory study \citep{mastrapa2006} where crystalline ice, if processed to an amorphous state via energetic particle bombardment, retains the spectral properties of its crystalline form. The properties of Nereid's surface may present an ideal opportunity to investigate such forms of water ice. 

\end{itemize}

\acknowledgements
This work was supported by a NASA Earth and Space Science Fellowship (PI: Sharkey) and Near-Earth Object Observations (NEOO) program grant NNXAL06G (PI: Reddy). We thank the IRTF telescope operators and MKSS staff for their support. The authors wish to recognize and acknowledge the significant cultural role and reverence the summit of Maunakea has always had within the indigenous Hawaiian community. We are most fortunate to have the opportunity to conduct observations from this mountain.

We wish to thank the two anonymous reviewers for providing additional context and thoughtful feedback which substantially improved this work.

\bibliography{sharkey_nereid_bib}{}
\bibliographystyle{aasjournal}

\begin{deluxetable*}{|c|c|c|c|c|c|}
\tablecaption{Spectral band parameters of detected absorption features derived without continuum removal.\label{tab:bandpars}}
\tablehead{\multicolumn{2}{|l|}{1.5 Micron Band Complex} & \multicolumn{2}{l|}{1.65 Micron Band} & \multicolumn{2}{l|}{2.0 Micron Band}} 
\startdata
Minimum ($\mu m$)      & Depth (\%) & Minimum ($\mu m$)      & Depth  (\%)  & Minimum ($\mu m$)      & Depth                 \\ \hline
$1.55_{-0.02}^{+0.02}$ & $13\pm2$   & $1.65_{-0.01}^{+0.02}$ & $10 \pm 3$   & $2.01_{-0.02}^{+0.01}$ & $27 \pm 3$             \\ \hline
\enddata
\end{deluxetable*}

\begin{deluxetable*}{c|c|c|c|c|c}
\tablecaption{Summary of model fit quality using $\chi_{R}^{2}$, and spectra band parameters derived without continuum removal.\label{tab:model_comps}}
\tablehead{\multicolumn{2}{c}{}\vline & \multicolumn{4}{c}{{\textit{Models}}}
} 

\startdata
 & & Cryst. Ice & Cryst. Ice & Cryst. Ice & Cryst. Ice \\
Band & Measured & Amorph. Ice & Magnetite & Amorph. Ice & Amorph. Ice \\
Parameter & Values & Magnetite & Amorph. Carbon & Tagish Lake & Amorph. Carbon \\\hline
1.5 $\mu m$ Band & $\mathbf{13 \pm 2}$& 0.17 & 0.14 & 0.13 & 0.08 \\
Depth (\%) & & & & & \\\hline
1.65 $\mu m$ Band & $\mathbf{10 \pm 3}$ & 0.09 & 0.10 & 0.07 & 0.03 \\
Depth (\%) & & & & & \\\hline
2.0 $\mu m$ Band & $\mathbf{27 \pm 3}$ & 0.28 & 0.23 & 0.23 & 0.18 \\
Depth (\%) & & & & & \\\hline
1.5 $\mu m$ Band & $\mathbf{1.55^{+0.02}_{-0.02}}$ & 1.52 & 1.54 & 1.52 & 1.53 \\
Minimum ($\mu m$) & & & & & \\\hline
1.65 $\mu m$ Band & $\mathbf{1.65^{+0.02}_{-0.01}}$ & 1.66 & 1.66 & 1.66 & 1.66 \\
Minimum ($\mu m$) & & & & & \\\hline
2.0 $\mu m$ Band & $\mathbf{2.01^{+0.01}_{-0.02}}$ & 2.01 & 2.03 & 2.01 & 2.02 \\
Minimum ($\mu m$) & & & & & \\\hline
$\chi_{R}^{2}$ & \nodata & 1.21 & 1.36 & 1.55 & 1.67 \\\hline
\enddata
\tablecomments{The model with crystalline and amorphous ice combined with magnetite returns the lowest $\chi_{R}^{2}$ by closely matching the 2.0-$\mu m$ band parameters at the expense of the 1.5-$\mu m$ band. The model with crystalline ice combined with magnetite and amorphous carbon returns a slightly higher $\chi_{R}^{2}$ by more closely matching the parameters of the 1.5-$\mu m$ band at the expense of the 2.0 $\mu m$ band. The other two models return poorer fits overall by failing to match either the observed continuum (the Tagish Lake model) or by failing to match any absorption band parameters (the amorphous carbon model).}
\end{deluxetable*}

\end{document}